\documentclass[preprint,aps]{revtex4}
\begin{document}

\title{Lattice-gas Monte Carlo study of adsorption in pores}
\author{Raluca A. Trasca, M. Mercedes Calbi, Milton W.Cole and Jose L. Riccardo $^*$}
\address{Department of Physics, Pennsylvania State University, University Park, PA 16802\\
$^*$ Departamento de Fisica, Universidad Nacional de San Luis, 5700 San Luis, Argentina} 

\begin{abstract}

A lattice gas model of adsorption inside cylindrical pores is evaluated with Monte 
Carlo simulations. The model incorporates two kinds of site: (a line of) ``axial'' sites
and surrounding ``cylindrical shell'' sites, in the ratio 1:7. The adsorption 
isotherms are calculated in either the grand canonical or canonical ensembles. At low
temperature, there occur quasi-transitions that would be genuine thermodynamic 
transitions in mean-field theory. Comparisons between the exact and mean-field theory
results for the heat capacity and adsorption isotherms are provided.
\end{abstract}

\maketitle

\section{Introduction}

One of the most exciting fields in condensed matter physics is the study of gases 
inside porous media \cite{gelb, chandler}. Its importance stems from questions of fundamental physics 
(e.g., dimensional crossover and the role of disorder) and a number of relevant 
technologies (e.g. catalysis, gas separation and storage) that utilize porosity. 
In some systems, the pores are fully interconnected so that a gas atom entering at
one point will eventually diffuse throughout the porous domain. In other cases of 
interest, individual pores are distinct, so that the problem can be thought of as 
essentially independent pores, with perhaps weak interpore interactions. Many model 
calculations of this adsorption have been presented. A large fraction of these 
consider the adsorption domain to consist of independent pores, either for 
simplicity or because that represents an accurate description of the geometry.
                            
In a recent paper (denoted I), we posited a particularly simple model of a nanoporous 
environment \cite{cnano}. That is a lattice-gas model with two kinds of site. One was a 
one-dimensional (1D) line of sites, which we call ``axial'' sites. Surrounding each 
axial site is a set of 7 ``cylindrical shell'' sites. The number 7 is chosen as an estimate
of the ratio of shell to axial densities for atoms of diameter $\sim 3.5 \mathring{A}$
in a carbon  nanotube of radius 7 $\mathring{A}$.
Each shell site has four 
shell neighbors and one axial neighbor. In the model, there are four interaction 
energies. $V_a$ is 
the potential energy of an axial atom due to its interaction with the host material, $V_s$ is that of a shell atom, $-\epsilon$ is the 
interaction energy between adjacent occupied sites of the same type, where $\epsilon>0$; 
$\epsilon_{sa}$ is the interaction between axial and shell sites, which could have either 
sign. Thus, the Hamiltonian is
\begin{equation}
H= N_a V_a + N_s V_s + H_{int}
\end{equation}
Here $N_a (N_s)$ is the number of occupied axial (shell) sites and the term $H_{int}$ involves 
both nearest neighbor interactions of the same species (axial-axial and shell-shell) and 
the axial-shell interactions, with their respective couplings.

In I, the adsorption isotherms were evaluated with mean-field theory (MFT). That is, 
the number of adsorbed particles (henceforth called atoms) was computed for specified 
reduced temperature $T^*=k_B T/\epsilon$ as a function of the reduced chemical potential 
$\mu^*=\mu/\epsilon$.
The resulting behavior found in I 
includes a set of transitions associated with filling the respective sites. If 
$V_a << V_s$, the axial sites are occupied first (as $\mu$ increases), while the reverse is 
the case if $V_a >> V_s$. If, instead, the energies are similar, there arises a 
``cooperative transition'', in which both sites are filled simultaneously at low T 
and sufficiently high $\mu$. These transitions, however, are artifacts because true 
thermodynamic transitions cannot exist in 1D. Nevertheless, the behavior in the exact 
solution is here found to be very similar to that of MFT, in that the coverage rises very rapidly near 
a threshold value of $\mu$. We note, however, that if there exists a transverse coupling between 
sites in neighboring pores such a genuine transition does  
occur (at a relatively high T - usually much higher than might be expected from
the strength of the the interpore 
interaction) \cite{cnano,book1,fisher,swift,gubbins,cole1}.

In this paper, we consider the same lattice gas model as that treated in I.  
The difference is that we here evaluate the system's properties with the Monte Carlo (MC) simulation 
technique. MC simulations of the lattice gas models have been
employed in studying a variety of adsorption and diffusion problems \cite{gelb,chandler,cole1,jose1,jose2}.
Our method is discussed in Section II along with a test of its sensitivity 
to the assumption of periodic boundary conditions. Section III presents our results and 
Section IV summarizes and comments upon them.

\section{Monte Carlo Method}

In the thermodynamic limit, a nanotube is a 1D system from the perspective of phase
transition theory (only one length 
approaches infinity). Thus, it is expected that gases adsorbed
in isolated nanotubes should behave thermodynamically like 1D systems. In Monte Carlo
simulations, one represents the system with a unit cell of sites that is repeated periodically.
In order to test the accuracy of the
simulations for various periodic cell sizes, we first perform Grand Canonical Monte Carlo (GCMC)
calculations for a purely 
1D line of sites. Fig.1(a) shows a comparison between isotherms obtained with GCMC 
(for different cell sizes) from MFT, and
the exact solution for an infinite Ising lattice gas \cite{pathria}. Note first that a spurious
singularity (an infinite slope) appears in the MFT curve, while no singularity is present
in the exact and MC results.
GCMC yields results very close to the exact results for cells consisting of 3 sites, 
replicated with periodic boundary conditions. 
We conclude that one does not need
large cell sizes to simulate isotherms for such a 1D system. A similar conclusion was found
by Swift {\it et al} when modeling adsorption in a porous medium \cite{swift}. 
As seen in Fig.1(b), specific heat results
obtained from Canonical Monte Carlo simulations are similar to 
the exact ones (at half occupancy) for cells of about 50 sites (25 particles). The variation
with number of sites in the cell is 
consistent with the thermodynamic relation:
\begin{equation}
\int_0^\infty C_N (T)/N dT= (E(\infty) - E(0))/N
\end{equation}
At half occupancy, 
the energy per particle at $T=\infty$ is the same for any cell size; 
but at $T=0$, it depends on the number of particles present in one cell ($N$) because the 
ground state of the 
periodic system at half occupancy consists of periodic islands of occupied sites: 
$ E(0)=-\epsilon (N-1)$. Thus the heat capacity increases with N as seen in Fig.1.

To make contact with our previous MF calculations for the system involving
axial and shell sites, we take the unit cell of the system to consist of one axial and seven
shell sites. 
The cell replicated periodically in simulations
is ten lattice constants long, meaning eighty sites in total.
We perform simulations in the grand canonical ensemble to find the evolution
of $N$ with $\mu$, and simulations in the 
canonical ensemble to find the specific heat $C_N(T)$ when the total
(axial + shell) number of particles is fixed. The specific heat is obtained from 
energy fluctuations according to the formula:
\begin{equation}
C_N/(N k_B)=(<E^2> - <E>^2)/(k_B T)^2
\end{equation}
Note that, in the canonical ensemble, even though $N$ is fixed, 
the axial and shell densities vary with T as
particles migrate from one shell to the other. This transfer process makes an interesting
contribution to
the specific heat, as described in the next section.

\section{Results}

Figures 2a, 2b and 2c compare results from MF and exact calculations of adsorption 
isotherms in three cases which differ in either the relation between $V_a$ and $V_s$ 
or the sign of the axial-shell interaction, $\epsilon_{as}$. The case illustrated in
$Fig. 2a$ is one for which the shell phase is energetically  favored relative to the 
axial phase ($V_s < V_a$) and the axial-shell interaction is attractive 
($\epsilon_{as} < 0$). At $T^*=0.5$, the shell fills in a nearly 
discontinuous way in the exact calculation; this might be called a ``quasitransition''
because of that behavior. The MFT, in contrast, exhibits behavior characteristic of a 
first-order transition (discontinuity) at $T^*=0.5$. In fact, the exact solution is 
numerically quite close to that of the MFT, so that a transition might be 
(incorrectly) inferred from experimental data that looks like this. At the higher 
value, $T^*=1$, the MFT shows critical behavior (a divergent slope at half-occupancy 
of the shell sites) while the exact result shows a smoother shell-filling behavior.
In Fig. 2a, the exact results show the axial phase formation to be gradual at $T^*=0.5$,
while the MFT behavior is that of a critical transition, since $T^*=z*J/2=2*\epsilon/4=
0.5$ is the 
critical temperature of the axial phase transition (where $z$ is the effective
coordination number, 2 for this transition). Under these circumstances, the 
shell phase provides a spectator field, which affects the critical value of $\mu$ but 
not the critical temperature in the MFT.

Fig. 2b displays rather different behavior of the isotherms, a consequence of the 
fact that $V_a$ and $V_s$ are very similar ($V_a$ = 20 and $V_s$ = 18). 
As a result, as discussed in I, there occurs a cooperative transition, in which both 
axial and shell sites fill together at the quasitransition. Moreover, 
because of the higher coordination number (including axial-shell attractions) in 
this case, the MFT critical temperature is pushed to a higher value than in the 
case (discussed above) of very different values of $V_a$ and $V_s$. This feature of 
the MFT is shared with the quasitransition of the exact solution. Evidence for this 
statement is seen in the similar steepness of the isotherm at $T^* = 1.2$ in Fig. 2b and that at 
$T^*=1.0$ in Fig. 2a; both have $10-90 \%$ widths $\Delta \mu ^* \simeq 1$.

Fig. 2c presents results that may seem counterintuitive at first sight. This behavior is 
a consequence of a repulsive axial-shell interaction, with $V_a < V_s$.
In this case, the axial phase forms at a low value of $\mu^*$, followed at higher $\mu^*$ 
by the appearance of the shell phase. The arrival of the shell phase, however, drives 
out the axial phase because of their mutual repulsion, so that the net increase in 
N is the difference between shell and axial occupancies (6=7-1). Eventually, at even higher $\mu$,
the axial phase finally returns to the pore. This behavior is precisely what is 
predicted in I with the MFT, as is seen in figure 2c. This represents a situation where the 
two phases don't ``fit'' particularly comfortably in the pore, but sufficient 
incentive, provided by $\mu$, can induce their coexistence. A sociological analogy 
to this behavior is a ``marriage of convenience''.

Figure 3 displays the specific heat  $C_N(T)$ under the circumstances corresponding 
to Fig. 2a, i.e. $V_s < V_a$ and an attractive axial-shell interaction. One observes 
two bumps in the data, near $T^*=0.3$ and $T^*=2.9$, respectively.
At low $T^*$, all of
the particles occupy shell sites (not filling them completely). The $T^*=0.3$ peak is 
associated with the loss of ordering among these particles; it would be a 
discontinuity in MFT (due to reaching the coexistence curve for the transition found 
in that model). The origin of the high $T^*$ (broad peak) behavior can be appreciated
from a comparison in Fig. 3 between $C_N (T)$ and $(E_s - E_a)d N_a/d T$. 
Here, the energy difference ($E_s -E_a$) equals the site energy difference $V_s -V_a$, 
plus a small correction due to the mutual interactions in the shell and axial phases.
Quantitatively, the peak region is described by the expected relation based on this interpretation:
\begin{equation}
C_N^{trans}(T^*)/(N k_B) = (E_s^* -E_a^*) (dN_a/dT^*)
\end{equation}
The broad peaks in both curves 
have maxima near $T^*=3$. This similarity indicates that the peak is a kind of 
desorption peak, familiar in film adsorption data. The difference here is that 
``desorption'' means a transfer of particles from the the lower energy shell to the 
axial phase as T increases. This happens, as expected, when $k_B$T is of order the 
site's energy difference, $V_s -V_a$. The axial-shell transfer heat capacity peak is analogous
also to the peak found in a recent study of adsorption on the outside of a bundle,
attributed to the transfer of molecules from the groove to the quasi- 2D surface of the
tubes \cite{mercedes}.  

Figs. 4a and 4b compare the behavior for two examples that differ in the size of the 
axial-shell energy difference, $V_s -V_a$. One set of curves corresponds to the case 
of a large difference, just discussed, featuring the high $T^*$ peak. The other set 
describes the cooperative quasitransition case, where the difference is small. The 
latter has its ordering peak at a higher T than the former, as discussed earlier; in 
addition, there is no particle transfer peak, as expected.

Fig. 5 compares two situations differing in the sign of the axial-shell interaction. 
As expected, the transfer peak occurs at a lower $T^*$ when this interaction is 
attractive than when it is repulsive; the reason is that the axial particles have a 
lower energy in the attractive case, so that the required excitation energy is smaller
than in the repulsive case.

Finally, in fig. 6 we explore the effects of varying shell occupancy fraction, at 
fixed interaction strength. A big difference between the curves appears in the low $T^*$ 
peak. For shell filling fraction 2/3, the peak occurs near $T^*=0.4$, lower than the 
other fillings' peak value, $T^*=0.6$. This difference can be understood from the MFT 
predictions. In that case, the in-shell transfer term is due to evaporation 
from the condensed to the dilute phase. This process stops when the condensed phase 
is evaporated completely, i.e. when the system reaches the coexistence curve of the 
in-shell transition. The temperature at which that occurs is lower at 2/3 filling 
than near 1/2; the other curves shown are at 1/2 and 5/12, respectively. The other 
notable feature in Fig. 6 is that the transfer peak occurs at lower $T^*$ in the 2/3 
filling case. This is probably due to the fact that the axial phase particles have a 
lower energy in this case because of the attractive axial-shell interaction energy, 
which is larger in magnitude at high occupancy than at low occupancy.

\section{Summary and conclusions}

In this paper, we have presented results from MC simulations of a number of cases 
involving different energies and interactions. A common feature is that properties 
computed ``exactly'' bear a close resemblance to those obtained from the MFT predictions.
This finding might be surprising in view of the fact that the system is 
essentially one-dimensional, meaning that the transitions in MFT are spurious. 
Nevertheless, as found previously in other systems \cite{swift}, the MFT yields very 
reasonable predictions away from the transition points. This finding suggests the 
broad utility of the MFT, a convenient situation because of its simplicity. From the 
experimental point of view, the difference between the two approaches may not even be 
visible. One should not, therefore, be surprised to see MFT-like behavior in such 
experimental results.

In closing, we note an obvious limitation of the lattice gas model- its inflexibility. 
Particles can occupy only a set of predetermined sites prescribed by the model. This means 
that a user of the model should think carefully about the choices of site 
and interaction energies. As indicated in I, a wide variety of behaviors can be found 
that depend on this set of parameters. Presumably, this reflects the variety seen in 
the many porous systems we are trying to describe.

This research is supported by NSF. We are grateful to Silvina Gatica for helpful comments.

\section {Figure captions}

1. Fractional site occupancy at $T^*=0.5$ for a purely 1D array of sites: (a) Occupancy as a function of 
reduced chemical potential. The exact analytic solution (full curve) is 
compared to GCMC
simulations with various simulation cell sizes (1 and 3 sites), and to the MFT results
(dotted curve). 
(b) Specific heat as a function of reduced temperature.
Results of canonical MC simulations with various numbers of sites (at half occupancy)
are compared to the exact solution for an infinite system at half occupancy.

2. Mean Field isotherms are compared to GCMC isotherms for three
cases: a) shell and axial energies differ appreciably ($V_s=20, V_a=12.5$)
and axial-shell interaction is attractive ($\epsilon_{sa} < 0$) b) shell and 
axial energies are
similar ($V_s=20, V_a=18$) and axial-shell interaction is attractive ($\epsilon_{sa} < 0$);
and c) shell and axial energies differ by an intermediate amount
($V_s=20, V_a=24$) and axial-shell interaction is repulsive ($\epsilon_{sa} > 0$)

3. Specific heat (full curve, left scale) and the transfer heat capacity (dashed curve,
right scale)
from shell to axial sites as a function of reduced temperature $T^*$, obtained
with the parameter set (a) of Fig. 2

4. ``Cooperative'' (full curve, $V_s = 20, V_a = 19$) vs. ``normal'' (dashed curve, 
$V_s = 20, V_a = 10$) behavior 
in  a) specific heat calculations, and b) axial and shell densities

5. Specific heat comparison between attractive ($\epsilon_{sa} < 0$) and repulsive 
($\epsilon_{sa} > 0$) axial-shell interaction. The parameters used are: $V_s=20,
V_a=10, N_p/N_s=1/2$ (where $N_p$ and $N_s$ are the number of particles and sites,
respectively).

6. Specific heat results for various number of particles: $ N_p/N_s=1/2$ (full
curve), $ N_p/N_s=5/12$ (dotted curve), and $N_p/N_s=2/3$ (dashed
curve). The parameters used are: $V_s=20, V_a=10$.

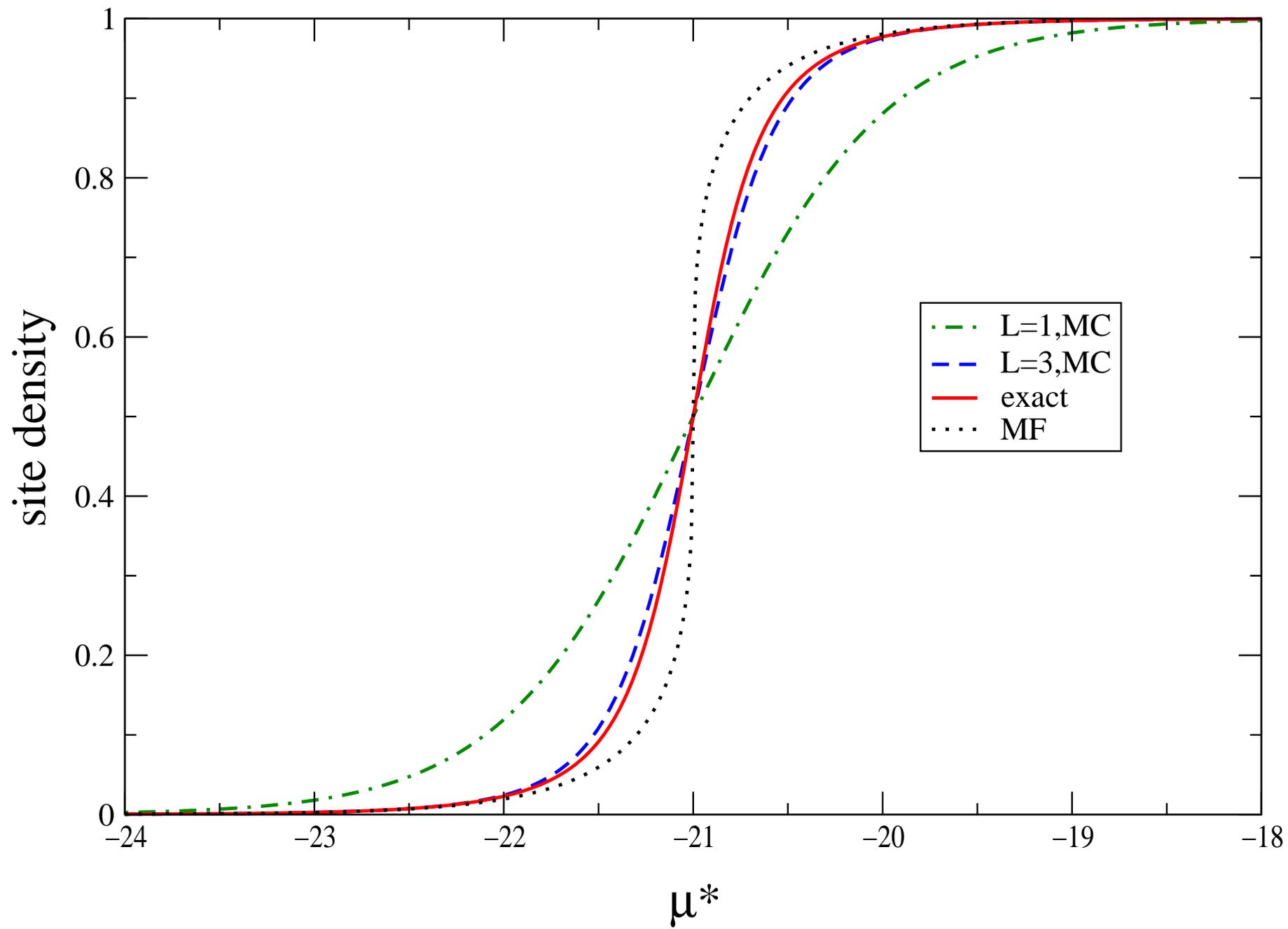
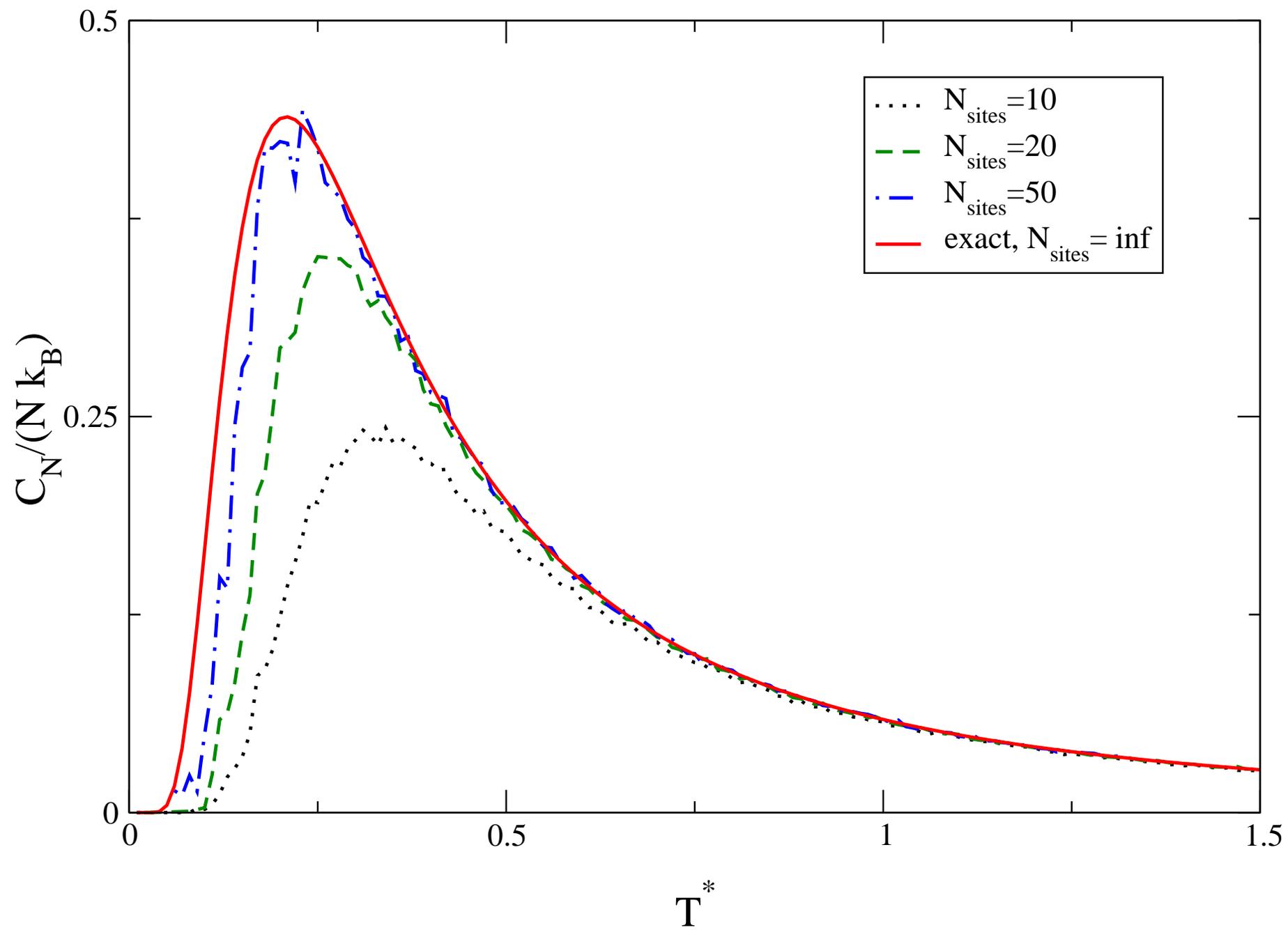
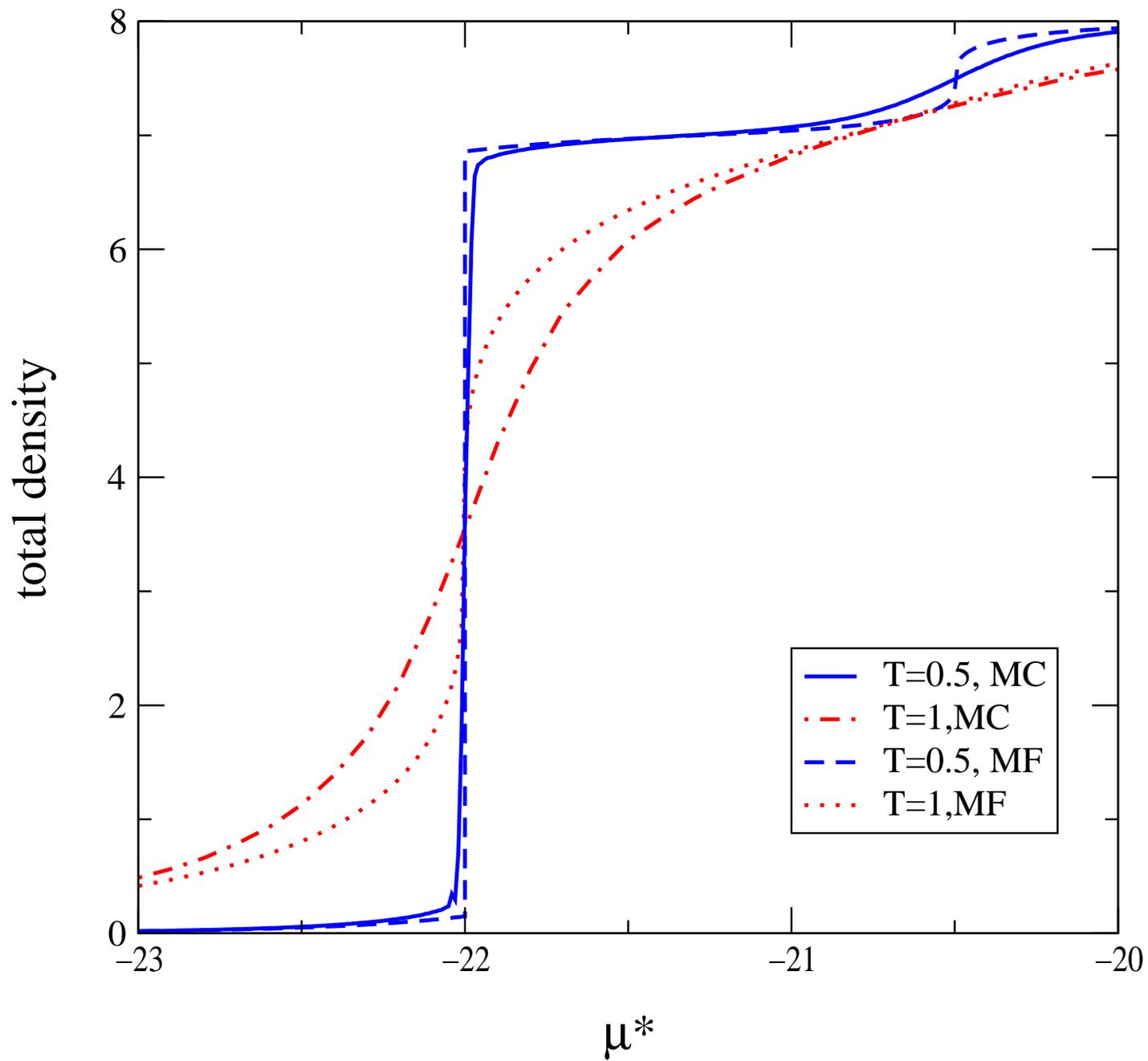
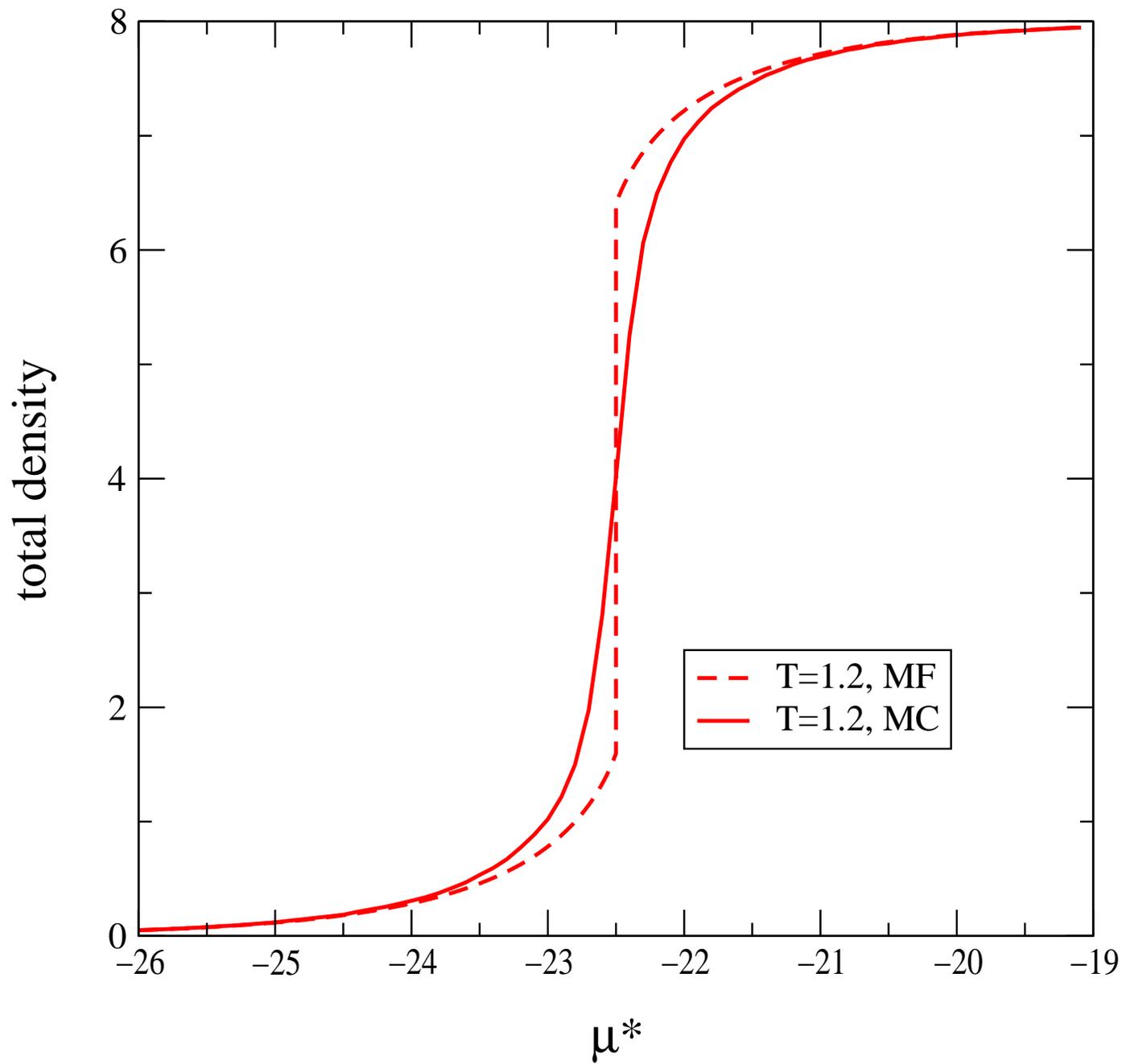
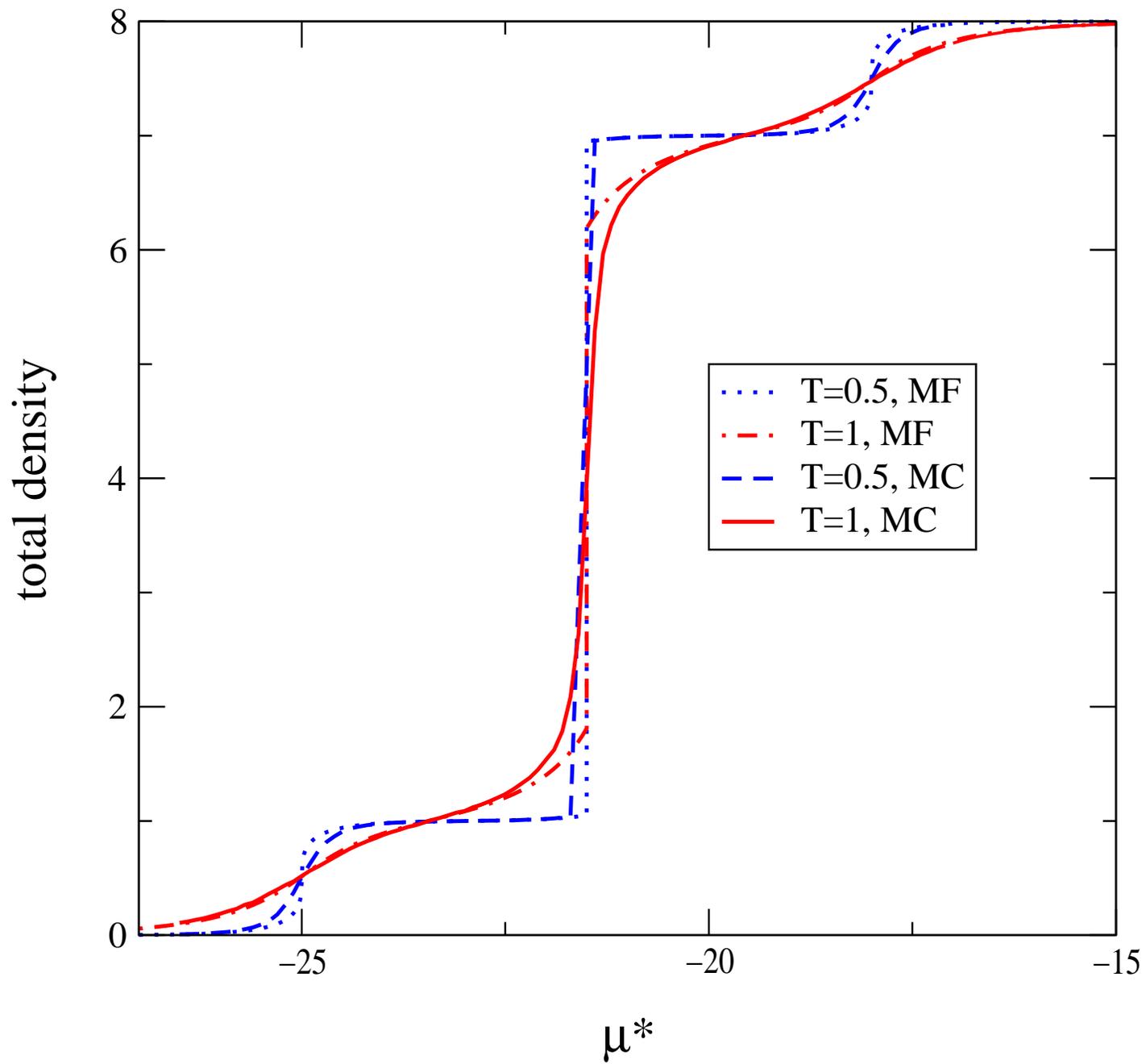
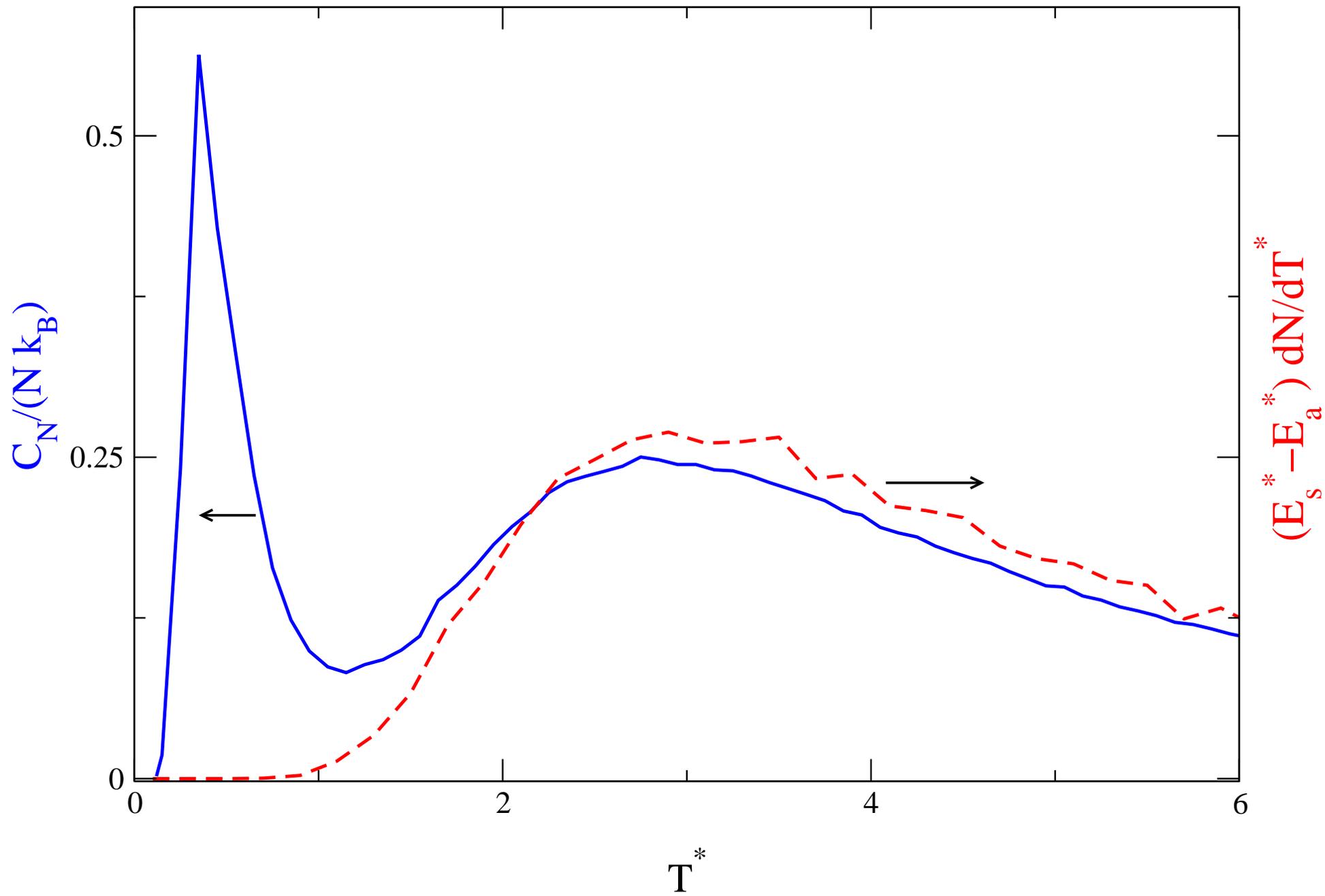
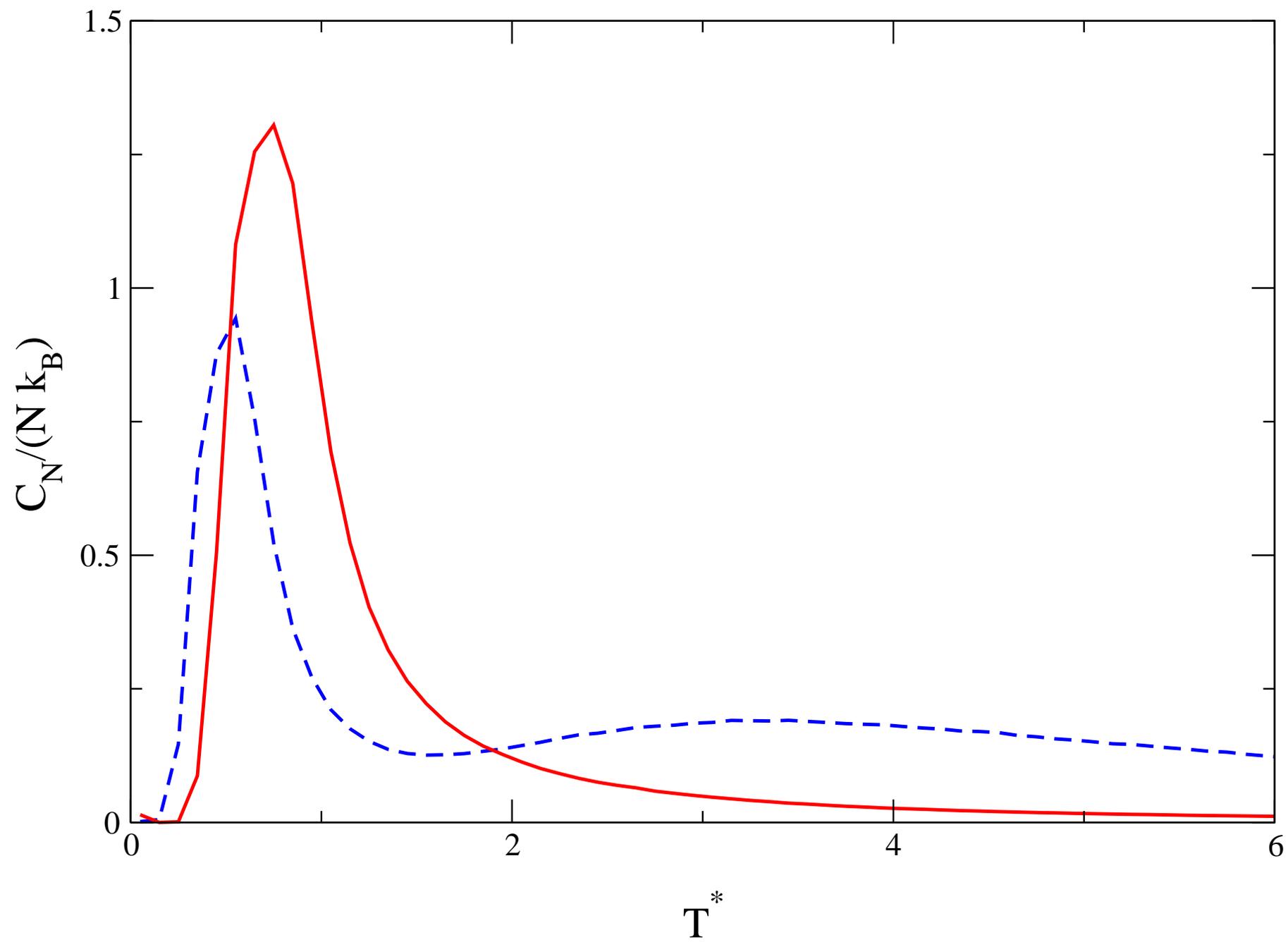
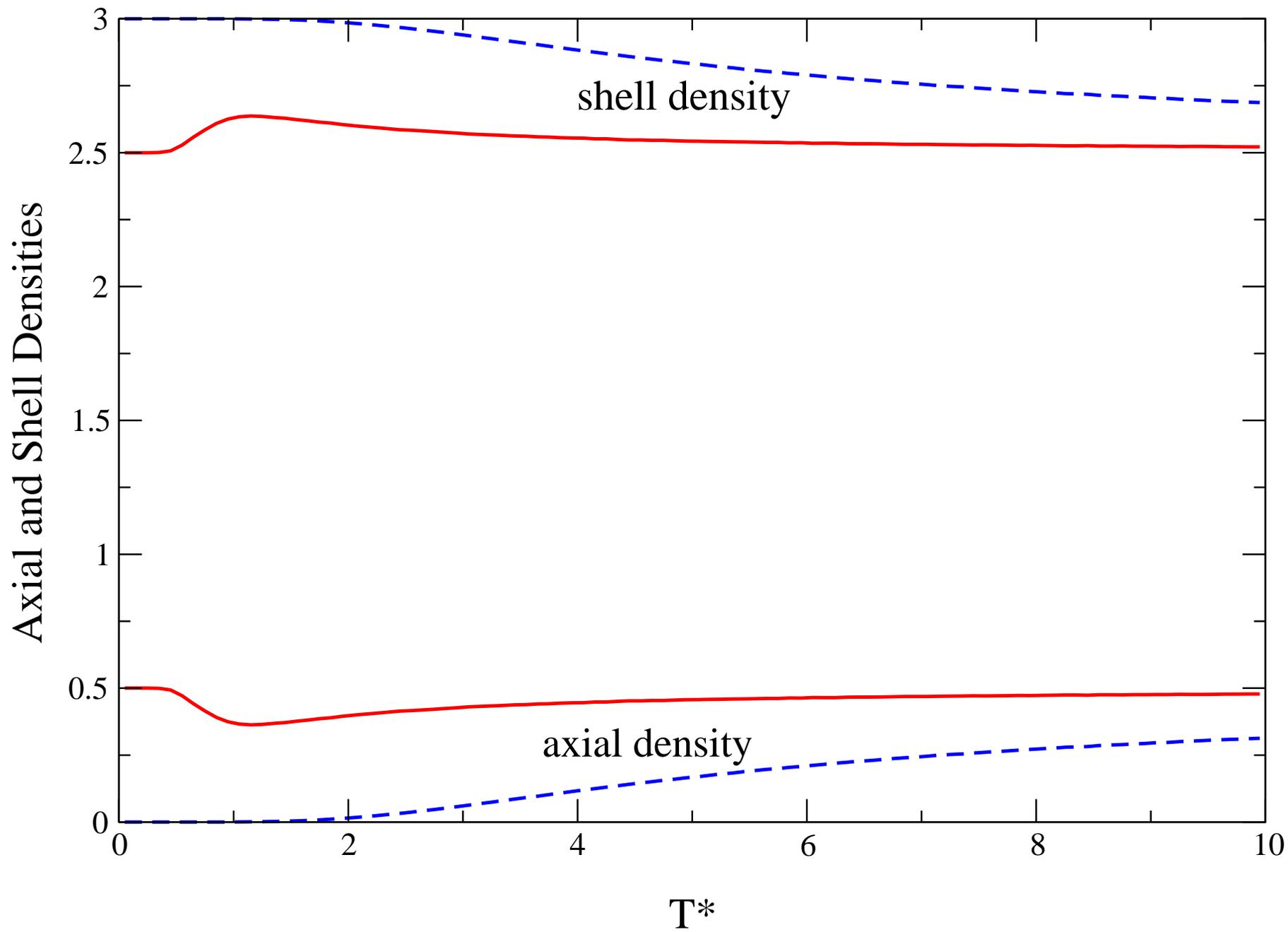
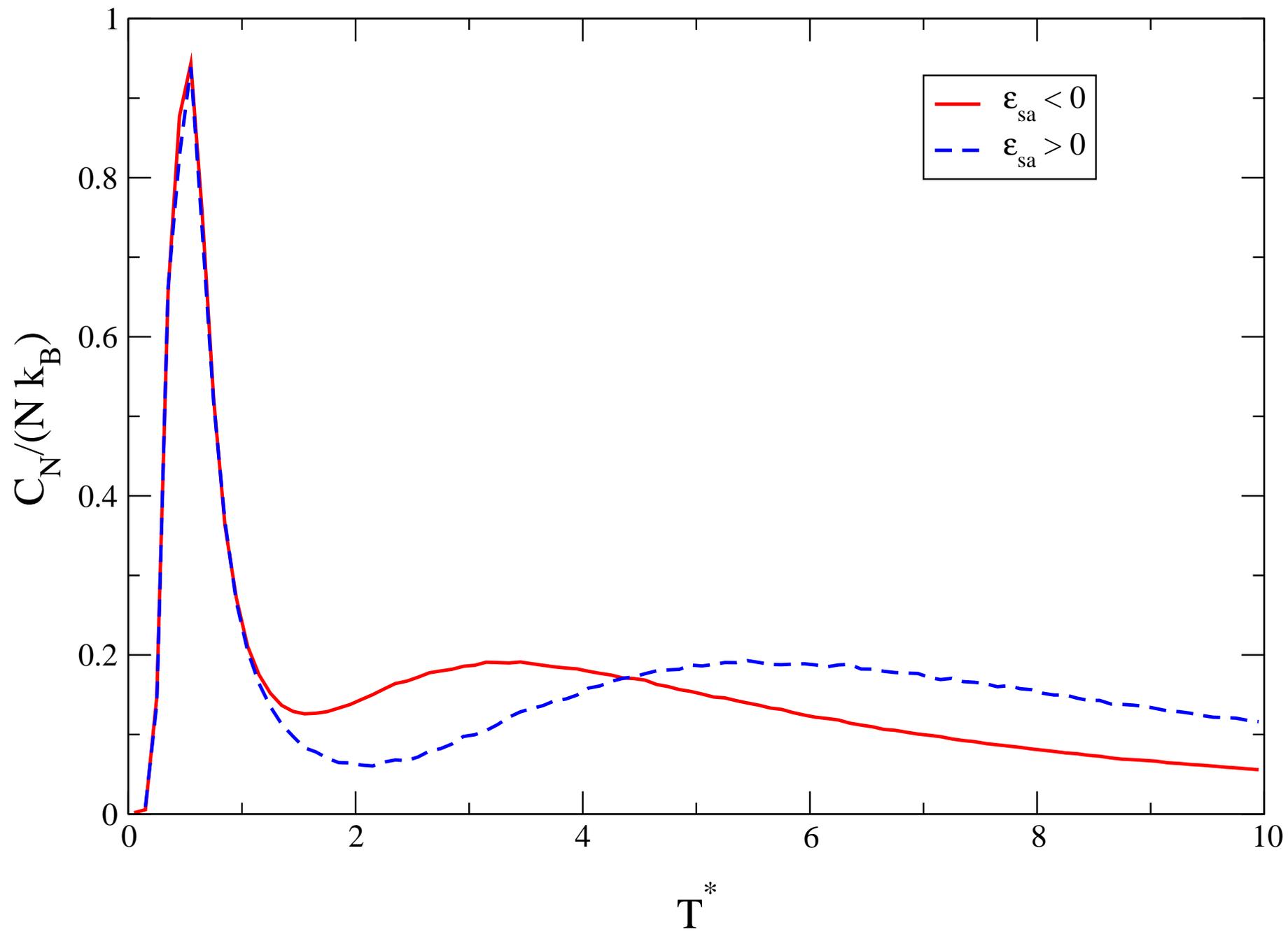
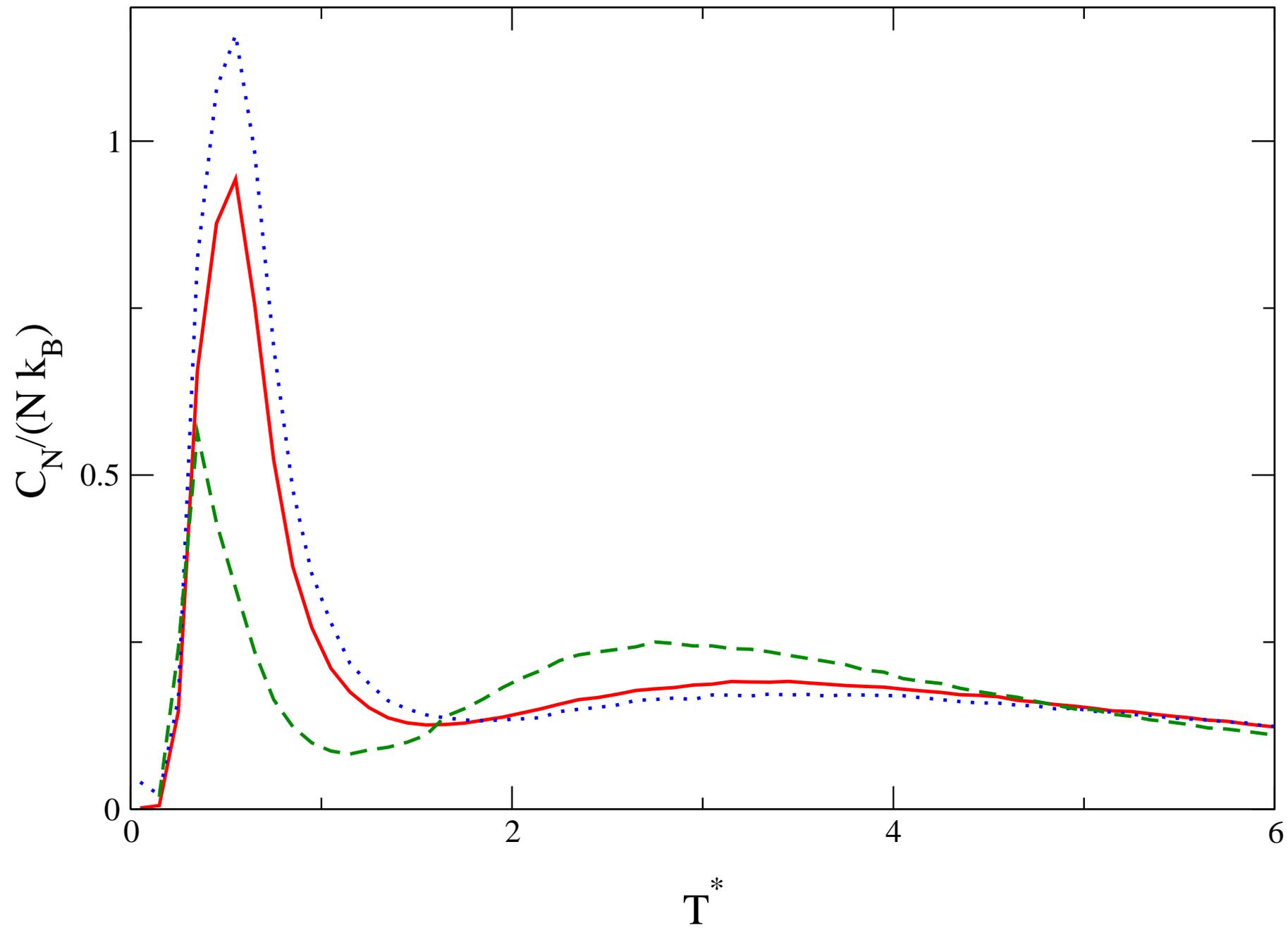
\end{document}